\documentclass[pre,twocolumn]{revtex4}
\usepackage{psfrag}
\usepackage{amssymb,amsmath,amsthm}
\usepackage[dvips]{graphicx}
\usepackage{verbatim}
\usepackage[colorlinks=true,linkcolor=blue,citecolor=blue]{hyperref}
\usepackage{subfigure}

\begin{document}

\title{Descending from infinity: convergence of tailed distributions}
%v1.2
\author{Christian Van den Broeck}
\affiliation{Hasselt University, B-3500 Hasselt, Belgium}
\author{Upendra Harbola}
\affiliation{Inorganic and Physical Chemistry, Indian Institute of Science,
Bangalore, India}
\author{Raul Toral}
\affiliation{IFISC (Instituto de F\'isica Interdisciplinar y Sistemas Complejos), Universitat de les Illes Balears-CSIC, Palma de Mallorca, Spain}
\author{Katja Lindenberg}
\affiliation{Department of Chemistry and Biochemistry and BioCircuits Institute,
University of California San Diego, La Jolla, CA 92093-0340, USA}

\begin{abstract}
We investigate the relaxation of long-tailed distributions under stochastic
dynamics that do not support such tails.  Linear relaxation is found to be a
borderline case in which long tails are exponentially suppressed in time but not
eliminated.  Relaxation stronger than linear suppresses long tails immediately,
but may lead to strong transient peaks in the probability distribution. 
A delta function initial distribution under stronger than linear decay 
displays not one but two different regimes of diffusive spreading. 
\end{abstract}


\date{\today}

\maketitle
\section{Introduction}

Since the discussion about the ``St Petersburg paradox" by the Bernouillis in
the early seventeen-hundreds about the fair fee required to play
a game with infinite average,  the study of probability distributions with long
tails or diverging moments has fascinated both scientists and non-scientists.
More recently,  stochastic 
processes giving rise to such long-tailed distributions have received a great deal of
attention \cite{Levy}.  The main purpose of this paper is to
investigate how distributions which initially have long or fat tails evolve
under stochastic dynamics that do not support such tails.  As we will see,
traditional (overdamped) linear Langevin
dynamics turn out to be a very interesting borderline case, exhibiting
sustained long tails which however decay exponentially in time. 
Langevin equation dynamics with decay rates stronger than linear instantly destroy fat tails,
but these may show up as transient maxima in the
probability distribution as it relaxes to its steady state form.

In the process of relaxing to the steady state, we observe another interesting phenomenon that we have not seen discussed in the literature.  While the usual expectation is that a delta function initial condition spreads diffusively until it reaches the steady state form, this turns out to be the case only in the case of linear dynamics.  When the decay rates are stronger than linear, the relaxation to the steady state displays not one but two distinctly separate regimes of diffusive spreading. We examine the origin of this phenomenon and determine the time at which the relaxation process transitions from one to the other.

To arrive at an understanding of the stochastic dynamics, we begin by analyzing the deterministic dynamics of equations of the form $dx/dt=-\gamma x^\alpha$ where $\alpha \geq 1$. These dynamics, simple as they are, already exhibit the underlying reasons for the unusual stochastic relaxation. We dedicate Sec.~\ref{DeterministicDynamics} to this analysis for delta function initial conditions.  These results in turn already reflect the interesting behavior found, still with deterministic dynamics, when the initial condition is distributed.  This is covered in Sec.~\ref{DistributedInitialConditions}. In the next two sections we add noise to the system, first in the case of linear relaxation in Sec.~\ref{LinearRelaxationWithNoise} and then for nonlinear relaxation in Sec.~\ref{NonlinearRelaxationWithNoise}.  We conclude with a short summary in the final section.  Some mathematical details are relegated to an appendix.

\section{Deterministic dynamics}
\label{DeterministicDynamics}

A direct integration of the  deterministic evolution equation
\begin{eqnarray}
 \label{eq-dd1}
\frac{dx}{dt}=-\gamma x^{\alpha}
\end{eqnarray}
leads to $x_t^{1-\alpha}=x_0^{1-\alpha}+(\alpha-1)\gamma t$, or
\begin{eqnarray}
 \label{eq-dd2}
x_t=\frac{x_0}{\{1+(\alpha-1)x_0^{\alpha-1}\gamma t\}^{\frac{1}{\alpha-1}}},
\end{eqnarray}
where $x_0$ is the initial condition. 
In the limit $\alpha\rightarrow 1$, Eq.~(\ref{eq-dd2}) reduces to the familiar
exponential
solution $x_t=x_0 e^{-\gamma t}$.
Without loss of generality we set $\gamma=1$, since this can always be achieved by rescaling the time variable, $t\to \gamma t$.
The solution (\ref{eq-dd2}) is a well-defined real function for all values of $x$ regardless of the value of $\alpha$ if the initial condition is positive, $x_0>0$.  If the initial condition is negative, $x_0<0$, then the requirement that $x^{\alpha-1}$ be a well-defined real function for all values of $x$ is satisfied, for instance, if $\alpha$ is an integer. Alternatively, we could replace Eq.~(\ref{eq-dd1}) with $\dot x=-\gamma x|x|^{\alpha-1}$ to remove the requirement that $\alpha$ must be an integer. All subsequent formulas are then valid for all $x$ if we replace $x^{\alpha-1}$ by $|x|^{\alpha-1}$.

We point out the following peculiar features of the above solution, see for example~\cite{strogatz}. 
For $\alpha>1$, the decay rate $x^{\alpha}$ becomes very strong for $x$ large,
so much so that ``infinity moves down" to a finite value 
$x^{+}_t$ at any finite time $t$. More precisely, the entire positive x-axis $x \in [0,\infty)$ is, for
any finite time $t$, mapped by the dynamics into a finite interval $[0,x^+_t)$,
with 
\begin{eqnarray}
 \label{eq-dd3}
x^{+}_t= \frac{1}{\{(\alpha-1)t\}^{\frac{1}{\alpha-1}}}.
\end{eqnarray}
This value is obtained by considering the limit $x_0\rightarrow \infty$ in
Eq.~(\ref{eq-dd2}). Note that one can rewrite Eq.~\eqref{eq-dd2} in the following more compact form:
\begin{eqnarray}
 \label{eq-dd21}
\frac{x_t}{x_0}=\left\{1+(x^+_t/x_0)^{1-\alpha}\right\}^{\frac{1}{1-\alpha}}.
\end{eqnarray}
On the other hand, for $0<\alpha<1$, the decay rate $x^{\alpha}$ of Eq.~\eqref{eq-dd1} remains significant for small $x$, so much so that all initial values smaller than a threshold value
$x^{-}_t$ will hit zero in a finite time $t$.
More precisely, the interval  $x \in [0,x^-_t)$ is mapped by the dynamics into
$0$ in the finite time $t$, with
\begin{eqnarray}
 \label{eq-dd3b}
x^{-}_t= \{(1-\alpha)t\}^{\frac{1}{1-\alpha}}.
\end{eqnarray}
This value is obtained by finding the value $x_0$ for which the denominator of Eq.~(\ref{eq-dd2}) vanishes,
$1+(\alpha-1)x_0^{\alpha-1}t=0$.

We mention in passing that one finds related opposite phenomena, i.e., reaching
infinity or escaping zero in a finite time, by considering $y=1/x$ with
$\dot{y}=y^{2-\alpha}$, with the understanding that the solution to such an equation
is only unique if the speed $\dot{y}$ has no singularity at the initial point~\cite{strogatz}. 

In the following, we focus on Eq.~(\ref{eq-dd1}) with
$\alpha \geq 1$.  We will illustrate several results for the particular choice
$\alpha=3$. In this case one has:
\begin{eqnarray}
\label{eq-dd4}
x_t=\frac{x_0}{\sqrt{1+2tx_0^{2}}}, \qquad
x^+_t=\frac{1}{\sqrt{2t}}.
\end{eqnarray}
These results are valid for all real values of $x$, with $x \in (-\infty,+\infty)$
mapped by the dynamics into the interval $(-x^+_t,+x^+_t)$. 
%%%%%%%%%%%%%%%%%%%%%%%%%%%%%%%%%%%

\section{Distributed initial conditions}
\label{DistributedInitialConditions}

The dynamics (\ref{eq-dd1}) with $\alpha > 1$ maps all the ``large" initial
conditions to the ``neighborhood" just below $x^+_t$. This raises the question
as to what happens when the initial probability distribution has a fat tail,
i.e., carries a significant probability weight for large $x$-values.
Let $P_0(x)$ denote the distribution of the initial conditions $x_0$. The
probability distribution $P_t(x)$ for the resulting $x$-values at time $t$ is
obtained from  the conservation of probability upon transformation of variables
(that is, from  $x_0$ to $x=x_t$):
\begin{eqnarray}
\label{eq-dic1}
P_t(x_t)=P_0(x_0)\left|\frac{dx_0}{dx_t}\right|.
\end{eqnarray}
By solving for $x_0(x_t)$ and calculating the derivative $\displaystyle\frac{dx_0}{dx_t}=\left(\frac{x_0}{x_t}\right)^\alpha$ from ({\ref{eq-dd1}), one thus finds:
\begin{widetext}
\begin{eqnarray}
\label{eq-dic2}
P_t(x) =P_0\left(x{\{1-\left(\frac{x}{x_t^+}\right)^{\alpha-1}\}^{\frac{1}{\alpha-1}}}\right)
 \left[1-\left(\frac{x}{x^+_t}\right)^{\alpha-1}\right]^{\frac{\alpha}{1-\alpha}}
\end{eqnarray}
\end{widetext}
for $x\in(-x^+_t,+x^+_t)$, and $P_t(x)=0$ otherwise.
 
To study the possible accumulation of probability for $P_t(x)$ in the
vicinity of $x^{+}$, we consider the following fat tail:
\begin{eqnarray}
\label{eq-dic5}
P_0(x)\sim x^{-\beta}.
\end{eqnarray}
One finds from Eq.~(\ref{eq-dic2}) for $x$ smaller than, but close to, $x^+_t$
\begin{eqnarray}
\label{eq-dic6}
P_t(x)\sim 
x^{-\beta}
\left[1-\left(x/x^+_t\right)^{\alpha-1}\right]^{\frac{\alpha-\beta}{
\alpha-1}}
\end{eqnarray}
We conclude that the distribution $P_t(x)$ has a divergence for $x \rightarrow
x^+_t$ for a sufficiently strong fat tail, i.e.  when $\beta < \alpha$. The
divergence is normalizable since $\beta>1$ in order for $P_0$ to be
normalizable. For $\beta=\alpha$, $P_t(x)$ converges to a nonzero value for $x
\rightarrow x^+_t$, while $P_t(x^+_t)=0$ for $\beta < \alpha$. 

As an interesting particular case, we focus on Lorentzian initial conditions,
\begin{eqnarray}
\label{eq-dic3}
P_0(x)=\frac{\lambda}{\pi}\frac{1}{\lambda^2+x^2}.
\end{eqnarray}
One finds from Eq.~(\ref{eq-dic2})
\begin{eqnarray}
\label{eq-dic4}
P_t(x)=\frac{\lambda}{\pi
}\frac{\{{1+(1-\alpha)tx^{\alpha-1}}\}^{\frac{\alpha}{1-\alpha}}}{\lambda^2+x^2\{{1+(1-\alpha)tx^{\alpha-1}}\}^{\frac{2}{1-\alpha}}}.
\end{eqnarray}
for  $x\in (-x_t^+,x_t^+)$, and $P_t(x)=0$ otherwise.
In particular, one has for $\alpha=3$ (see also Fig. \ref{divergence}):
\begin{widetext}
\begin{eqnarray}
\label{eq-dic5b}
P_t(x)=\frac{\lambda}{\pi
}\frac{1}{\sqrt{1-2tx^2}\left(x^2+\lambda^2(1-2tx^2)\right)}, \qquad x\in(-1/\sqrt{2t},1/\sqrt{2t}).
\end{eqnarray}
\end{widetext}
Note the divergences at the endpoint of the interval
$[-1/\sqrt{2t},1/\sqrt{2t})$. In particular,
\begin{eqnarray}
\label{eq-dic7}
P_t(x)\sim_{x\rightarrow 1^-/\sqrt{2t}} \frac{\sqrt{2}\;t}{\pi \;\sqrt{1-{x}\sqrt{2t}}}.
\end{eqnarray}
%%%%%%%%%%%%%%%%%%%%%%%%%%%%%%%%%%%
\section{Linear relaxation with noise}
\label{LinearRelaxationWithNoise}

Our main purpose is to study the relaxation in the presence of additive noise.
No
exact analytic results are available for the case of nonlinear relaxation, hence
we first
 turn to the study of linear relaxation with the exponent $\alpha=1$. 
As we show below, this  case  can be studied in  analytic detail, with the
additional bonus that it is an interesting and revealing borderline case, in
particular with respect to the persistence of the long tails. 
We  consider the following linear Langevin equation:
\begin{eqnarray}
\label{eq-l1}
\frac{dx}{dt}=- \gamma x+\xi,
\end{eqnarray}
with $\xi$ Gaussian white noise with mean value and correlations:
\begin{eqnarray}
\label{eq-l2}
\langle\xi(t)\rangle&=&0,\\
\langle\xi(t)\xi(t^\prime)\rangle&=&2D \delta(t-t^\prime).
\end{eqnarray}
In the following, we again set $\gamma=1$ by a suitable rescaling of the time variable $t\to \gamma t$ and the noise intensity $D\to D/\gamma$. One could also scale out the noise intensity (i.e. set $D=1$ by a redefinition of variables provided $D>0$), but we keep the $D$ dependence in order to reproduce the noiseless limit $D=0$ discussed in the previous section. 
The equivalent Fokker-Planck equation reads:
\begin{eqnarray}
\label{eq-l3}
\frac{\partial P_t(x)}{\partial t} =\frac{\partial }{\partial x}
(x\;P_t(x))\;+\;D\frac{\partial^2 }{\partial x^2} P_t(x).
\end{eqnarray}
The exact solution for the probability distribution $P_t(x)$, starting from a
delta distribution $P_0(x)=\delta(x-x_0)$, is a Gaussian with first two
central moments 
\begin{eqnarray}
\label{eq-l4}
\mu_t&=&\langle x\rangle_t=x_0 e^{-t},\\\nonumber\\
\sigma^2_t&=&\langle (\delta x)^2\rangle_t=D (1-e^{-2t}).
\label{eq-l4b}
\end{eqnarray}
For a general initial condition $P_0(x)$ one thus finds:
\begin{eqnarray}
\label{eq-l5}
 P_t(x) = \int dx_0 \frac{ e^{ \frac{-\left( x-\mu_t\right)^2}{2
\sigma^2_t } } }{\sigma_t\sqrt{2\pi }} P_0(x_0).
\end{eqnarray}
We now introduce the Fourier transform, $\hat{P}_t(k)=\int_{-\infty}^{\infty}dx\,e^{ikx}P_t(x)$, which coincides with the moment generating function:
\begin{eqnarray}
\label{eq-l6}
\langle e^{ikx}\rangle_t=\sum_{n=0}^{\infty}\frac{(ik)^n}{n!}\langle x^n \rangle_t,
\end{eqnarray}
when all moments exist.
One finds:
\begin{eqnarray}
\label{eq-l7}
\hat{P}_t(k) =e^{-\frac{1}{2} \sigma_t^2 k^2}\hat{P}_0(k\;e^{- t}),
\end{eqnarray}
where $\hat{P}_0(k)$ is the Fourier transform of the initial distribution
$P_0(x)$.
This result leads to the following general conclusion. Consider an initial
distribution with a long tail in the sense that some or all of its moments are
divergent. The divergence of moments is equivalent to the fact that the moment
generating function cannot be written as a Taylor expansion around $k=0$, i.e.,
it is a non-analytic function of  $k$ at $k=0$. 
According to Eq.~(\ref{eq-l7}), this non-analyticity will not be removed and in fact
will persist for all time while keeping the same character (same type of
non-analyticity). Nevertheless, the influence of the non-analyticity is
suppressed exponentially in time. We conclude that, while strictly speaking, any
type of long tail will persist in the same form for all finite times, its effect
will become very difficult to observe for times much longer than the decay time
as its weight is exponentially suppressed. 

To investigate the situation in  more detail, we turn to 
Lorentzian initial conditions, cf. Eq.~(\ref{eq-dic3}). From the known result
\begin{eqnarray}
\label{eq-l8}
\hat{P}_0(k)=e^{-\lambda|k|},
\end{eqnarray}
we get
\begin{eqnarray}
\label{eq-l9}
\hat{P}_t(k) &=& e^{-|k|\lambda_t-{\frac{1}{2} \sigma_t^2 k^2}},\\
\lambda_t&=&\lambda e^{-t}.
\end{eqnarray}
Transforming back to real space, one obtains after a simple manipulation,
\begin{eqnarray}
 \label{eq-l10a}
P_t(x)&=&\frac{1}{\pi}\int_{-\infty}^{+\infty} dk\,\hat{P}_t(k)\nonumber\\
&=&\frac{e^{-\mu_t^2/2\sigma_t^2}}{\pi}\int_{\mu_t/\sigma_t^2}^\infty dq\,\cos\left(x(q-\frac{\mu_t}{\sigma_t^2})\right)e^{-\frac12\sigma_t^2q^2}.\nonumber
\\
\end{eqnarray}
%which can alternatively be expressed as~\footnote{For this we use the tabulated integral
Using the tabulated integral
$$
\int_0^b dx\,\cos(2ax)e^{-x^2}=\frac{e^{-a^2}}{4}\sqrt{\pi}\left(\textrm{erf}(b+ia)+\textrm{erf}(b-ia)\right)
$$
as well as the property $\textrm{erf}({\overline{z}})=\overline{\textrm{erf}(z)}$, Eq.~\eqref{eq-l10a} can alternatively be expressed as
\begin{eqnarray}
 \label{eq-l10}
P_t(x)=\Re\left[\frac{e^{-(x-i\lambda_t)^2/2\sigma_t^2}}{\sigma_t\sqrt{2\pi}}
\textrm{erfc}\left(\frac{ix+\lambda_t}{\sigma_t\sqrt{2}}\right)
\right],
\end{eqnarray}
where $\textrm{erfc}(z)=1-\textrm{erf}(z)$ is the complementary error function. The time evolution of the probability distribution starting from an initial Lorenztian form to the final Gaussian shape can be observed in Fig.~\ref{convergence}.

In the limit of vanishing noise, $D\to 0$, this result reduces to the deterministic limit, cf. limit $\alpha\to 1^-$ of Eq.~(\ref{eq-dic4}):
\begin{eqnarray}
 \label{eq-3a}
\lim_{D\to 0} P_t(x) 
= \frac{1}{\pi} \frac{\lambda_t}{x^2+\lambda_t^2},
\end{eqnarray}
where we have used the following
asymptotic form of the error function, cf.~\cite{abramowitz}: 
\begin{eqnarray}\label{erfa}
\textrm{erfc}(z)\sim_{|z|\to\infty} \frac{e^{-z^2}}{z\sqrt{\pi}}.
\end{eqnarray}
This asymptotic form assumes $\textrm{arg}(z)<3\pi/4$, a condition satisfied by the argument of the error function in Eq.~(\ref{eq-l10}).

Turning to the long time limit $t\to\infty$, one finds that the distribution function Eq.~(\ref{eq-l10}) converges, as expected, to the Gaussian stationary
solution of Eq. (\ref{erfa}):
\begin{eqnarray}
 \label{eq-l12}
P^{st}(x)= \frac{e^{-\frac{x^2}{2D}}}{\sqrt{2\pi D}}.
\end{eqnarray}
However, the approach to this asymptotic result retains, at all finite times, the trace of the initial long-tailed distribution. Indeed, as already indicated via the analysis in Fourier space, cf. Eq.~(\ref{eq-l9}), the asymptotic decay of the distribution as $1/x^2$ for
$x\rightarrow \infty$ persists for all times, even though it is exponentially suppressed in time. This can be derived directly from the explicit expression Eq.~(\ref{eq-l10}) for the probability density, by again invoking Eq.~(\ref{eq-3a}) (see also Fig.~\ref{fig-tail}):
\begin{eqnarray}
P_t(x)\sim_{|x|\to\infty} \frac{\lambda e^{-t}}{\pi}x^{-2}.
\end{eqnarray}

\section{Nonlinear relaxation with noise}
\label{NonlinearRelaxationWithNoise}

We now turn to the investigation of the behavior of long tailed distributions under nonlinear relaxation with noise, 
\begin{eqnarray}
\label{eq-nl1}
\frac{dx}{dt}=-\gamma x^\alpha+\xi,
\end{eqnarray}
with equivalent Fokker-Planck equation
\begin{eqnarray}
\label{eq-FP}
\frac{\partial P_t(x)}{\partial t} =\frac{\partial }{\partial x}
(\gamma x^\alpha\;P_t(x))\;+\;D\frac{\partial^2 }{\partial x^2} P_t(x).
\end{eqnarray}
We first note that a simple dimensional analysis leads to the general scaling relation
\begin{widetext}
\begin{equation}
\label{eq-nl3}
P_t(x;D,\gamma)=\left(\frac{\gamma}{D}\right)^{\frac{1}{\alpha+1}}P_{\tau}(\zeta), \hspace{30pt}\zeta=\left(\frac{\gamma}{D}\right)^{\frac{1}{\alpha+1}}x,\hspace{10pt}\tau=\left(\gamma^2 D^{\alpha-1}\right)^{\frac{1}{\alpha+1}}t,
\end{equation}
\end{widetext}
which, without loss of generality, allows us to set $\gamma=D=1$ in the Fokker-Planck equation. 

An object of prime interest
is the Green function $G_t(x|x_0)$, i.e., the solution of Eq.~(\ref{eq-nl1}) for the initial condition $P_0(x)=\delta(x-x_0)$. The general solution of Eq.~(\ref{eq-nl1}) can then be written as:
\begin{eqnarray}
\label{eq-nl4}
 P_t(x) = \int dx_0 G_t(x|x_0) P_0(x_0).
\end{eqnarray}
When comparing with the linear case, cf. Eq.~(\ref{eq-l5}), two different difficulties are encountered in the  application of this result, for example to an initial Lorentzian distribution.  First, the explicit expression for $G_t(x|x_0)$ is not known. Second, and in our context more importantly, $G_t(x|x_0)$ does not have the simple dependence on $x - \textrm{constant}\times x_0$, which, in the linear case,  allowed us to write the above integral as a convolution. That led to the simple explicit expression Eq.~(\ref{eq-l7}) in Fourier space, with the immediate conclusion that initial long tails in the linear case survive for all times.  As we will see below, and as expected from the previous deterministic analysis, this is no longer the case for stronger than linear relaxation.

Before turning to a numerical solution of Eq.~(\ref{eq-FP}), we present a perturbative solution for the Green function, revealing a surprising feature about the interplay between nonlinear relaxation and noise.
We suppose that the stochastic trajectory $x$ starting at a given initial position $x_0$ can be well approximated by the deterministic trajectory $x_t$ starting at the same initial position, $x= x_t+\delta x$ with $\delta x$ small. 
This approximation is expected to be valid for short times.
The explicit form for $x_t$ is given in Eq.~(\ref{eq-dd2}).
The equation for $\delta x$ reads:
\begin{eqnarray}
\frac{\delta x}{dt}=-\alpha\; x_t^{\alpha-1} \delta x +\xi,
\end{eqnarray}
where we neglected terms of order $( \delta x)^2$. This approximation is expected to be valid when $\langle (\delta x)^2 \rangle_t\ll x_t^2$.
We conclude that $ \delta x$ is a Gaussian random variable, hence we need only evaluate its first two moments.
Since the initial condition for the deterministic trajectory is the same as the initial condition of the stochastic trajectory, we have that 
$\langle  \delta x \rangle_{t=0} =0$ and hence $\langle  \delta x \rangle_t =0$ at all times. For the second moment $\langle (\delta x)^2 \rangle_t$, we find:
\begin{eqnarray}\label{ss}
\frac{d\langle (\delta x)^2 \rangle_t}{dt}&=&-2\alpha\; x_t^{\alpha-1}\langle (\delta x)^2 \rangle_t +2\\
&=&-\frac{2\alpha}{(\alpha-1)t+x_0^{1-\alpha}}\langle (\delta x)^2 \rangle_t +2.
\end{eqnarray}
We conclude that the short time motion corresponds to the deterministic trajectory, onto which is superimposed a Brownian motion in a harmonic
well with spring constant softening as $1$/time. This has the following surprising consequence. The Green function is Gaussian in the short-time limit, but
displays two different diffusive regimes. Indeed, one finds from Eq.~(\ref{ss}) that
\begin{eqnarray}
\langle (\delta x)^2 \rangle_t=2 \int_0^t \; d\tau\;\left\{\frac{(\alpha-1)\tau+x_0^{1-\alpha}}{(\alpha-1)t+x_0^{1-\alpha}}\right\}^{\frac{2\alpha}{\alpha-1}},
\end{eqnarray}
where we used the fact that $ \langle (\delta x)^2 \rangle_{t=0}=0$.
At  very short times, the ``ballistic" deterministic dynamics ($\sim t$) is slow compared to the diffusion induced by the noise term ($\sim \sqrt{t}$), and we have  a usual diffusive regime:
\begin{eqnarray}
\langle (\delta x)^2 \rangle_t=2t\;\;\;  \mbox{for}\;\;\;x_0^{1-\alpha}\gg(\alpha-1)t,
\end{eqnarray}
cf. the similar expression in the short-time regime for linear relaxation, Eq.~(\ref{eq-l4b}). In the case of nonlinear relaxation, for instance $\alpha=3$, the time-regime in which this behavior can be observed is very small for $x_0>1$. 
For longer times (but still short enough such that $\langle (\delta x)^2 \rangle\ll x_t^2$), one however finds a second diffusive regime, but with suppressed diffusion coefficient:
\begin{eqnarray}
\label{eq-deltax2}
\langle (\delta x)^2 \rangle_t=\frac{\alpha-1}{3\alpha-1} 2t\;\;\;  \mbox{for}\;\;\;x_0^{1-\alpha}\ll(\alpha-1)t.
\end{eqnarray}
The suppression is by a factor $4$ for $\alpha=3$ and by a factor $2/7$ for $\alpha=5$. 
Note that the cross-over time between the two regimes is given by the condition $x_0=x_t^+$, that is, the cross-over time for a given $x_0$ is equal to the time needed for the deterministic dynamics to come down to $x_0$ from infinity, cf. Eq.~(\ref{eq-dd3}). In particular, the time diverges for the case of linear relaxation $\alpha=1$, and hence this second diffusive regime ceases to exist in that case.
 
One can use the short-time Gaussian form for the Green function to get an approximate solution for distributed initial conditions, namely:
\begin{eqnarray}
\label{eq-nl3b}
 P_t(x) \approx_{t\rightarrow 0} \int dx_0 \frac{ e^{ \frac{-\left( x-x_t\right)^2}{2
\sigma^2_t } } }{\sigma_t\sqrt{2\pi }} P_0(x_0),
\end{eqnarray}
where $x_t$ is the deterministic trajectory specified in Eq.(\ref{eq-dd2}) and $\sigma^2_t=\langle (\delta x)^2 \rangle_t$, as given in Eq.~(\ref{eq-deltax2}). A numerical analysis confirms that  this approximation is  quite good in this time regime for large $x_0$ and, therefore, correctly reproduces the short time behaviour of the tail of the  distribution. By changing
 variables $x_0\to x_t$ we can use the property $P_t^\textrm{det}(x_t)dx_t=P_0(x_0)dx_0$ where $P_t^\textrm{det}(x_t)$ is the deterministic pdf as given in Eq.~(\ref{eq-dic2}). Hence one can explicitly perform the Fourier transform of Eq.~(\ref{eq-nl3b}): 
\begin{equation}
\hat{P}_t(k)=e^{-\frac{\sigma_t^2}{2}k^2}\hat{P}_t^\textrm{det}(kx_t).
\end{equation}

It is difficult to obtain exact analytic results valid for all times, so we next turn to numerical simulations. We encountered numerical instabilities when using standard methods for simulating either the Langevin equation Eq.~(\ref{eq-nl1}) or the Fokker-Planck equation Eq.~(\ref{eq-FP}). We therefore developed an alternative numerical integration method, which is explained in some detail in the appendix, and which seems to be stable and reliable. 

First, we confirm the existence of the two different diffusive regimes. We numerically evaluate the Green function starting from the value $x_0=5$ for $\alpha=3$ and $\alpha=5$. We clearly identify three time regimes. In the first two time regimes, the Green function is Gaussian, but displays the above predicted switch-over from a $\langle (\delta x)^2\rangle_t\sim 2t$ to a $\langle (\delta x)^2\rangle_t\sim \frac{\alpha-1}{3\alpha-1}2t$ behavior. This is illustrated in more detail in Fig.~\ref{moments}, where we plot $\langle (\delta x)^2\rangle_t$ as a function of time. The third time regime corresponds to the relaxation to the steady state, with a saturation value
\begin{equation}
\label{asym}
\frac{\int_{-\infty}^{\infty}x^2e^{-x^4/4}}{\int_{-\infty}^{\infty}e^{-x^4/4}}=\frac{2\Gamma(3/4)}{\Gamma(1/4)}=0.675978\dots
\end{equation}
Second, in Fig.~\ref{pdf-FP} we reproduce the relaxation of an initial Lorentzian distribution in a potential with $\alpha=3$. Again, the numerical results are in full agreement with the analysis given above. We recall that the deterministic relaxation projects the entire real axis onto a finite interval $(-x^+_t,x^+_t)$, with normalizable divergences at the boundaries.
The effect of the additive noise is to wash out the divergences, leading to Gaussian peaks in the vicinity of $\pm x^+_t$, with diffusive spreading described by Eq.(\ref{eq-deltax2}), 
$\langle{(\delta x)^2\rangle_t}\sim t/2$.   Both peaks move in towards zero relatively slowly, as $1/\sqrt{t}$. This picture is valid for short to intermediate times. Note also the somewhat surprising non-monotonic behaviour in time of the probability density in the vicinity of $x=0$. Probability mass  first flows out of this region, with the density decreasing below the Lorentzian values. At a later time, the probability peaks generated by the deterministic dynamics from the tails of the initial distribution bring in probability mass towards the center region, and the probability density again increases to finally attain its  steady state value, which is above the Lorentzian value.

\section{Discussion}
\label{Discussion}

A large number of phenomena in science are described in terms of linear dynamics. Yet, linear relaxation, $\dot{x}=-x$ and the concomitant exponential dependence on time, $\exp(-t)$, describe 
borderline situations when compared to nonlinear dynamics $\dot{x}=-x^\alpha$, with $\alpha\neq1$. For example, if an initial condition includes contributions at $x\to\infty$, the exponential takes an infinitely long time to bring these contributions down from infinity. Any initial condition takes forever to reach $x=0$. In other words, any initial contribution that decreases with time takes an infinite time to reach the final condition. This is no longer the case when nonlinear relaxation is considered. Trajectories come down from infinity instantaneously  for an exponent $\alpha>1$, while trajectories corresponding to an  exponent $\alpha<1$  hit zero in a finite time.

In this paper, we showed that linear dynamics remains a borderline case in the presence of additive noise ($\dot{x}=-x^\alpha+\xi$ with $\xi$ Gaussian white noise). We focused on the comparison between linear relaxation $\alpha=1$ and nonlinear relaxation with $\alpha>1$.
We found that linear dynamics will sustain long tails for all times, if initially present, even though the weight of these tails is suppressed exponentially in time.
Nonlinear relaxation, however, will instantaneously kill any long tails. As an unexpected by-product of our analysis, we mention the discovery of a ``second diffusive regime" for noisy nonlinear dynamics. By this we mean the following.
The propagator (Green function) for the linear Langevin equation is exactly Gaussian. The average follows the exponential decay dictated by the deterministic dynamics. The variance $\sigma^2$ displays the expected short time diffusive behavior $\sigma^2=2Dt$, where $D$ is the noise intensity, followed by saturation towards the steady value for larger times. For nonlinear dynamics with additive noise, the propagator is still Gaussian in a short-time regime.
The average  again reproduces the (nonlinear) deterministic dynamics. The variance has an interesting behavior different from that of the linear problem. Apart from the ``usualÓ short time behavior $\sigma^2=2Dt$, which the nonlinear problem also exhibits, another regime of linear dispersion follows as time increases, but with reduced coefficient, i.e.,  $\sigma^2=2DÕt$ with $DÕ=D(\alpha-1)/(3\alpha-1)$. This second regime of ``suppressed diffusion" is actually the dominant regime before the saturation to the steady state, for initial conditions starting sufficiently far away from zero. The crossover time between the two regimes scales as $x_0^{1-\alpha}/(1-\alpha)$, which diverges as $\alpha\to 1$. Therefore, notably, the second regime is completely absent for linear dynamics.

%%%%%%%%%%%%%%%%%%%%%%%%%%%%%%%%%%%%%%%%%%%%%%%%%%%%

\begin{acknowledgments}
RT and CVdB acknowledge the warm hospitality at UCSD where this work was carried out. UH
acknowledges the support of the Indian Institute of Science, India. RT acknowledge financial support from EU (FEDER) and the Spanish MINECO under Grant INTENSE@COSYP (FIS2012-30634) and CVdB from MO 1209 COST action of the European Community. KL acknowledges the support of the National Science Foundation under Grant No. PHY-0855471.
\end{acknowledgments}

\section*{Appendix: Numerical integration of Eq.~(\ref{eq-FP})}\label{App-A}
For the numerical integration of Eq.~(\ref{eq-FP}) we have used a splitting method combining the exact solutions of the purely deterministic ($D=0$) and purely stochastic ($\gamma=0$) limits of the equation. They read respectively (we use the notation $P(x,t)$ for $P_t(x)$)
\begin{widetext}
\begin{eqnarray}
P(x,t+h)&=&(1+(1-\alpha)\gamma h x^{\alpha-1})^{\frac{\alpha}{1-\alpha}}P\left(x(1+(1-\alpha)\gamma h x^{\alpha-1})^{\frac{1}{1-\alpha}},t\right),\label{eq-app-det}\\
\hat{P}(k,t+h)&=&e^{-Dhk^2}\hat{P}(k,t)\label{eq-app-sto},
\end{eqnarray}
\end{widetext}
where we use the expression in Fourier space for the stochastic solution. In the numerical method we discretize space $x_i=idx,\, i\in[-M+1,M]$. Hence, $P_i(t)$ accounts for the probability in the whole interval $[x_i,x_{i+1})$. After setting the initial condition $P_i(t=0)$, the method works as follows:
\begin{enumerate}
\item
Given $x_i,\, i\in[-M+1,M]$, compute $x'_i=x_i(1+(1-\alpha)\gamma h x_i^{\alpha-1})^{\frac{1}{1-\alpha}}\equiv a_i x_i$. 
Find the index $i'=\textrm{floor}[x'_i/dx]$.
(The function $\textrm{floor}[z]$ is defined as the largest integer less than or equal to the real (positive or negative) number $z$). Implement 
Eq.~(\ref{eq-app-det}) using linear interpolation in the interval $[x_{i'},x_{i'+1})$, namely:
\begin{equation}
P'_i(t)=a_i^\alpha\left[(P_{i'+1}(t)-P_{i'}(t))\cdot(ia_i-i')+P_{i'}(t)\right].
\end{equation}
\item
2) Compute the Fourier transform $\hat{P}'_q(t)$ of $P'_i(t)$ with $q\in[-M+1,M]$. Apply Eq.~(\ref{eq-app-sto}) using
\begin{equation}
\hat{P}_q(t+h)=e^{-Dhk_q^2}\hat{P'}_q(t), \quad k_q=\frac{\pi}{Mdx}q.
\end{equation}
Invert the Fourier transform to find $P_i(t+h)$.
\end{enumerate}

Although this method is accurate only to order $O(h)$, we have found it more convenient than others which in principle have a higher order of precision, such as 2-nd order Runge-Kutta, as it can handle the stiffness of the deterministic part as well as implementing a very efficient pseudo-spectral algorithm for the stochastic part.

For the calculations of the Fourier transforms we have used fast Fourier routines. We typically take $M=2^{16}$ and $dx=10^{-3}$, so the interval value for $x$ is approximately $(-65.5,+65.5)$. Depending on initial conditions we use $h=10^{-3},\,10^{-4},\,10^{-5}$ and check in every case that results with smaller values of $h$ do not deviate significantly.

%%%%%%%%%%%%%%%%%%%%%%%%%%%%%%%%%%%%%%%%%%%%%%%%%%%%

\newpage

\begin{figure}[p]
\centering
\includegraphics[width=8cm]{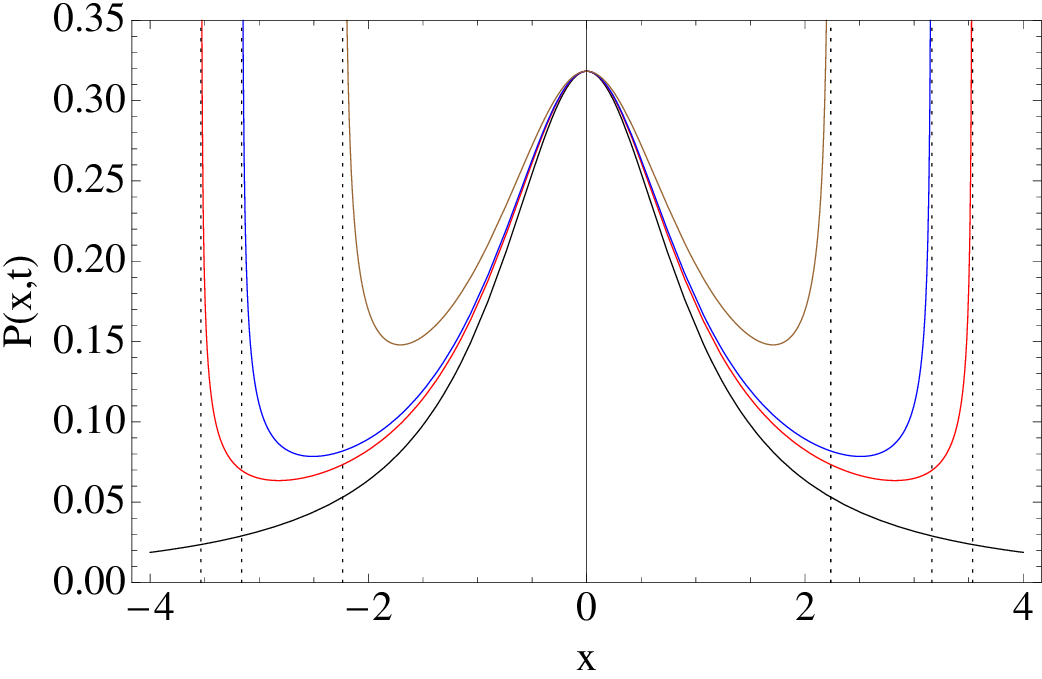}
\caption{(Color online) Time evolution of an initial Lorenzian distribution Eq.~(\ref{eq-dic3}) with $\lambda=1$
under the deterministic dynamics Eq.~(\ref{eq-dd1}) with $\alpha=3, \,\gamma=1$  ($t=0,\,0.04,\,0.05,\,0.10$). The initial distribution $t=0$ is shown as a black curve. Time increases from black to
brown ($t=0.10$).
Note the divergences of the probability
distribution at the endpoints of the interval $(-1/\sqrt{2t},1/\sqrt{2t})$.}
\label{divergence}
\end{figure}

\begin{figure}[p]
\centering
\includegraphics[width=8cm]{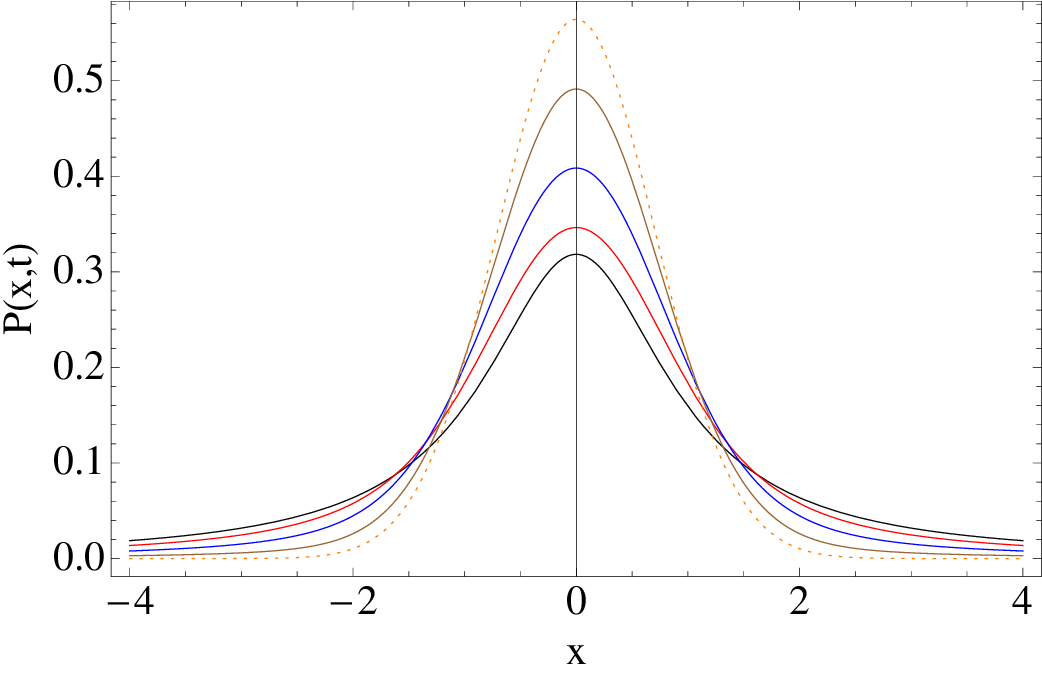}
\caption{(Color online) Convergence of the initial Lorenzian distribution to the Gaussian
distribution
under linear Langevin evolution. The initial distribution is shown as a black
curve. Time increases from
bottom to top. The dotted curve is the steady state Gaussian distribution.}
\label{convergence}
\end{figure}
%%%%%%%%%%%%%%%%%%%%%%%%%%%%%%%%%%%%%%%%%%%%%%%%%%%%
\begin{figure}[p]
\centering
\includegraphics[width=8cm]{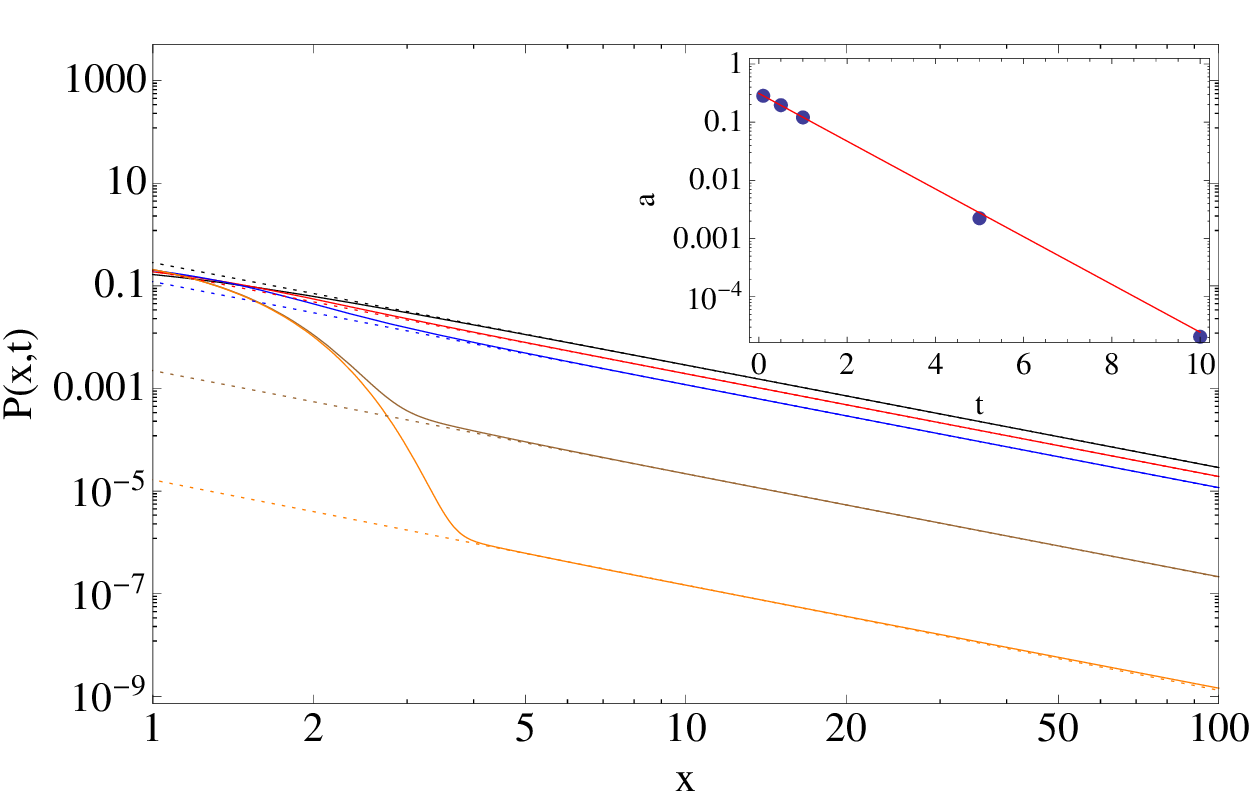}
\caption{(Color online) Tails of the distributions shown in  Fig.~\ref{convergence} on a log-log
plot. Dotted lines represent fits: $ax^{-2}$. The amplitude $a$ decreases exponentially
with time as shown in the inset.}
\label{fig-tail}
\end{figure}

\begin{figure}[p]
\includegraphics[width=18cm]{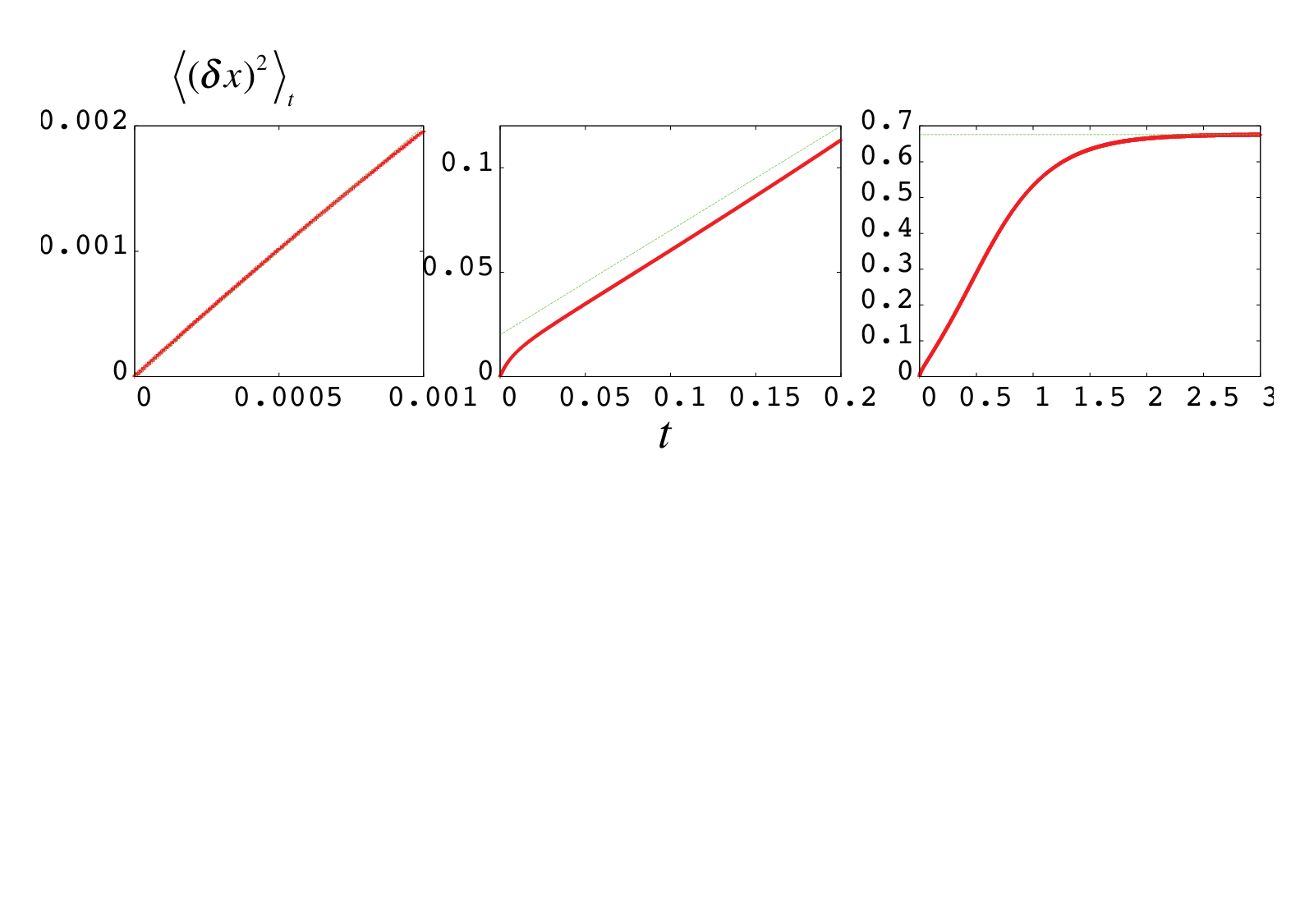}
\vspace{-2.5in}
\caption{(Color online) Plot of the dependence with time of the variance $\langle (\delta x)^2\rangle_t$ of the probability distribution $P_t(x)$ obtained from a numerical solution of the Fokker-Planck Eq.~(\ref{eq-FP}) in the non-linear case with $\alpha=3$. We have taken as initial condition a delta-function centered at $x_0=5$. The numerical results are given by the red curves. We clearly see the three regimes predicted by the theory: (top) early time with normal diffusion where the variance $\langle(\delta x)^2\rangle_t$ grows as $2t$ (green line), (middle) intermediate time with a reduced diffusion, variance growing as $t/2$ (green line), and late time saturation where the asymptotic value is given by Eq.~(\ref{asym}).
\label{moments}}
\end{figure}

\begin{figure}[p]
\centering
\includegraphics[width=14cm]{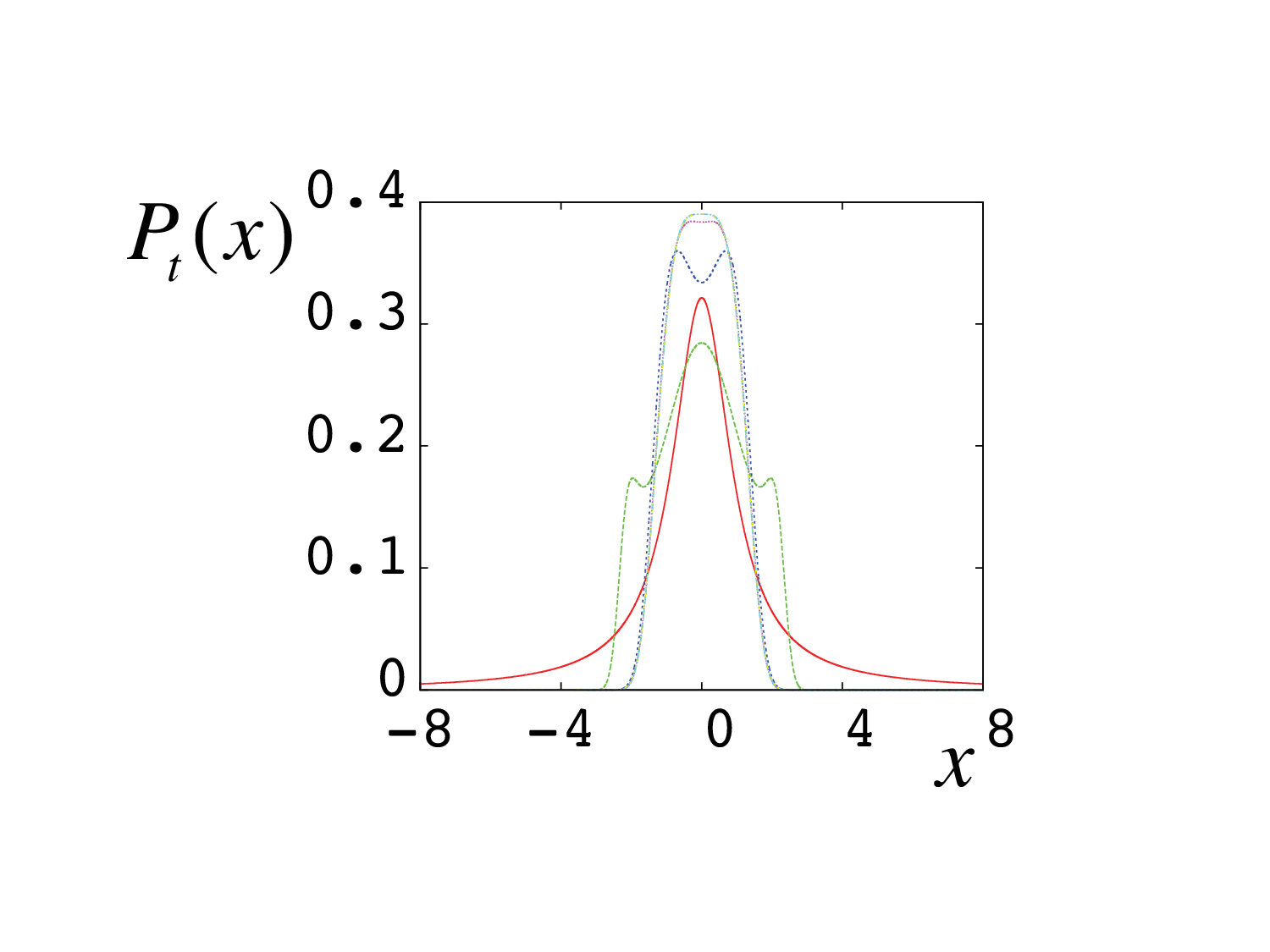}
\vspace{-1.0in}
\caption{(Color online) Probability distribution functions obtained from a numerical integration of the 
Fokker-Planck equation, Eq.(\ref{eq-FP}), for $\alpha=3$ and $\gamma=1,\, D=1$ using the method explained in the appendix. The initial condition is the Lorenztian distribution Eq.~(\ref{eq-dic3}) with $\lambda=1$. From outwards to inwards the curves correspond to $t=0,\,0.01,\,0.1,\,0.5,\,1,\,10$,  the last curve coinciding exactly with the stationary distribution $P_\textrm{st}(x)=\frac{\sqrt{2}}{\Gamma(1/4)}e^{-x^4/4}$.
\label{pdf-FP}}
\end{figure}

\end{document}